# Title: How attention simplifies mental representations for planning

Authors: Jason da Silva Castanheira[1], Chang (Christina) He[1], Nicholas Shea[2], Stephen M. Fleming[1,3,4,*]

Affiliations:
[1] Department of Experimental Psychology, University College London, 26 Bedford Way, London WC1H 0AP, UK
[2] Institute of Philosophy, School of Advanced Study, University of London, London, UK
[3] Max Planck UCL Centre for Computational Psychiatry and Ageing Research, University College London, London WC1B 5EH, UK
[4] Canadian Institute for Advanced Research (CIFAR), Brain, Mind and Consciousness Program, Toronto, ON M5G 1M1, Canada

[*] Corresponding author: j.castanheira@ucl.ac.uk

## Abstract:

Human planning is efficient—it frugally deploys limited cognitive resources to accomplish difficult tasks—and flexible—adapting to novel problems and environments. Computational approaches suggest that people construct simplified mental representations of their environment, balancing the complexity of a task representation with its utility. These models imply a nested optimisation in which planning shapes perception, and perception shapes planning – but the perceptual and attentional mechanisms governing how this interaction unfolds remain unknown. Here, we harness virtual maze navigation to characterise how spatial attention controls which aspects of a task representation enter subjective awareness and are available for planning. We find that spatial proximity governs which aspects of a maze are available for planning, and that when task-relevant information follows natural (lateralized) contours of attention, people can more easily construct simplified and useful maze representations. This influence of attention varies considerably across individuals, explaining differences in people's task representations and behaviour. Inspired by the 'spotlight of attention' analogy, we incorporate the effects of visuospatial attention into existing computational accounts of value-guided construal. Together, our work bridges computational perspectives on perception and decision-making to better understand how individuals represent their environments in aid of planning.

## Keywords
Planning, Attention, Computational model, Cognitive neuroscience, Psychology, Visuospatial attention, Computational psychology.

## Significance statement:

Humans have an impressive ability to plan. Theoretical models in computer science propose that instead of using all the available information in a scene, a decision-maker should form a simplified mental representation of their environment over which they plan. However, little is known about how perceptual and attentional processes shape the planning process in humans. We find that people form simplified mental representations in line with the natural contours of spatial attention, whereby information limited to a visual hemifield is more readily available for planning, like a spotlight illuminating a part of the environment. We develop a novel computational model of the effects of attention on planning and characterise systematic variation between individuals in how they simplify their mental representations.

## Introduction

Humans have an impressive ability to plan. We are able to model the world, simulate potential outcomes, and select among possible courses of action. Take, for example, your first trip to London. You want to visit Buckingham Palace despite being jet-lagged. Looking at a map of the underground, you're overwhelmed with information but need to make a plan. How do you solve this problem? Even simple decisions like this involve many potential actions and outcomes, making it impossible to systematically evaluate every possible option, especially given limited cognitive resources (Callaway et al., 2022; Daw et al., 2005; Dezfouli & Balleine, 2013; Griffiths et al., 2019; Huys et al., 2012; Newell & Simon, 1956). Explaining how people plan efficiently and flexibly under these constraints is a long-standing challenge in human and machine intelligence (Daw et al., 2005; Griffiths et al., 2019; Hassabis et al., 2017; Saxe et al., 2021).

Theories of human problem-solving conceptualize planning as a search through a 'decision tree' of all potential actions and their outcomes (Breslow & Aha, 1997; Callaway et al., 2022; Huys et al., 2012; Newell & Simon, 1956; Quinlan, 1986). In our example, an individual may first list all possible tube stations within walking distance and then evaluate which action sequence will get them closer to their destination. Previous work proposes different algorithmic strategies for how an agent efficiently searches over a complex decision tree. These strategies include ignoring low-value actions (i.e., pruning) (Huys et al., 2012, 2015; Knuth & Moore, 1975; Mingers, 1989), limiting how far in the future one might search (i.e., depth )(Callaway et al., 2022; Keramati et al., 2016; Korf, 1985; Snider et al., 2015), or relying on previously learnt strategies (i.e., habits) (Daw et al., 2005; Dezfouli & Balleine, 2013; Keramati et al., 2016; Kool et al., 2016).

This previous work, however, largely assumes that a decision-maker has a fixed representation of the problem. When planning involves constructing and evaluating multiple multi-step trajectories within a decision tree, the computational burden increases with the complexity of the representation of the problem space. Consider planning in a two-dimensional spatial grid, for example. A finer-grained grid presents many choice points about which way to turn. A coarser-grained grid presents fewer choice points.

Since the number of branches is a multiplicative function of the number of choice points, a simplified representation of the task space, if chosen appropriately, can have a profound effect on reducing the computational demands of planning.

One elegant approach to forming such a simplified representation is to adaptively select the granularity of information required to complete the task (Ho et al., 2022), known as value-guided construal (VGC). Unlike previous accounts, which model human planning as a search over all items (e.g.., tube lines), the VGC model predicts that a cognitively limited decision-maker selects a manageable subset of information over which to plan—i.e., a task representation—balancing utility and complexity (Ho et al., 2022). In our example, the VGC algorithm would plan over a few relevant tube lines rather than planning over all possible stations. To select the representation that achieves the best balance between utility and complexity, the model searches across all possible combinations of tube lines, computing the value (i.e., the plan's utility minus its cost) of each representation for planning a specific journey. The algorithm then selects the representation with the highest value, which ensures that an ideal observer selects a representation which only includes the items (i.e., tube lines) that lead to successful planning while excluding as many items as possible to keep the plan as simple as possible. For our purposes, items included in the representation are considered task-relevant, while items that are not represented are considered task-irrelevant. This algorithm, therefore, provides a normative standard of an efficient plan to which we can compare people's actual plans.

In previous work, Ho and colleagues discovered that people's awareness of, and memory for, obstacles in a maze varies in line with the predictions of a VGC model. The VGC model implies two nested optimisations – an outer loop of construal, and an inner loop that runs a plan conditional on a particular task representation. The VGC model is a normative model and remains agnostic as to the cognitive mechanisms controlling the construal. In particular, the perceptual and attentional mechanisms governing *how* information is selected to become part of a task representation remain unknown. Initiating such a nested computation plausibly rests on inductive biases – general principles that a perceptual system can apply to select task-relevant information, before refining it as part of the planning process (Gershman, 2021). Selective attention is proposed as one general mechanism by which the brain selects relevant information, either voluntarily (endogenous) or reflexively (exogenous)(Carrasco, 2011; Carrasco et al., 2004; Chica et al., 2013; Landry et al., 2021, 2023; Nobre & Kastner, 2014).

Previous studies have demonstrated that attention guides the selection of particular features of the environment to support reinforcement learning(Niv, 2019). However, it remains unknown whether and how attention shapes value-guided construal “on the fly” during planning. For instance, one possibility is that forming a simplified task representation is a “late” passive side-effect of the planning process – a tendency to focus on what we are thinking about. Alternatively, VGC may reflect an “early” selection of perceptual information, perhaps based on a rapid feedforward sweep of perceptual input(V. A. Lamme & Roelfsema, 2000). These alternatives echo classic debates between early and late selection models of attention(Broadbent, 1958; Nelson et al., 2012), but now situated within the broader landscape of computational accounts of planning. More generally, despite the wealth of literature on attention, and pioneering efforts to incorporate attentional constraints into models of decision-making(Ho et al., 2022; Niv, 2019), we lack a basic understanding of how attention influences planning.

To make progress on this question, we examined the role of visuospatial attention on how people construct simplified task representations across three experiments in human participants. We build on previous work using maze navigation to provide a rich readout of people’s current task representations. We predicted that if visuospatial attention is guiding the formation of task representations, the construal process will be constrained by inductive biases characteristic of attentional selection. For instance, previous work has illustrated how attentional selection is biased by the spatial context in which information is presented: presenting distractors alongside task-relevant stimuli makes attentional selection more challenging(He et al., 1996; Liu et al., 2009; Whitney & Levi, 2011). Attention, in this case, spills over to the neighbouring stimuli. These findings align with the metaphorical attentional spotlight or zoom lens, which stipulates that the focus of visual attention can move around the visual field, illuminating a limited spatial extent at a time (Norman, 1968; M. Posner et al., 1980). According to this model, individuals can, for example, orient their attention preferentially to a single hemifield—i.e., lateralizing—which is enabled by a hemispheric lateralization of alpha power over posterior cortex(Bagherzadeh et al., 2020; Jensen, 2024; Keefe & Störmer, 2021; Landry et al., 2024).

Our focus in this study was to examine how participants perceive and represent their environment (the maze stimulus). This is a distinct process to how participants orient their attention during navigation itself, which is not part of our current study.  To do so, we harness classical signatures of attentional selection to characterise how visuospatial attention shapes awareness of maze obstacles during planning. First, we demonstrate “attentional overspill”: participants preferentially incorporate task-irrelevant information into their task representation when it is presented in spatial proximity to task-relevant information. Second, we observe that attentional overspill is reduced when task-relevant information is lateralised to a single hemifield, allowing participants to more effectively form optimal task representations. Finally, we extend the VGC model to incorporate visuospatial attention as a key psychological mechanism for constructing simplified task representations. Together, our findings furnish a computational account of how attention and perception guide simplified representations in the service of planning.

## Results

To examine the role of visuospatial attention in planning, we relied on a previously developed maze navigation paradigm in which participants solved 2-D mazes (Figures S1-6) (Ho et al., 2022), avoiding obstacles obstructing their path (Figure 1a, left panel). We operationalized planning using a maze navigation paradigm, akin to our tube-related example, where participants were required to plan a route through the maze, avoiding obstacles that blocked their path. Obstacles predicted by the sVGC model to be included in the representation were considered task-relevant.

At the end of every trial, participants reported their awareness of specific obstacles (see Methods for details). The level of awareness reported for different obstacles provides a read-out of what features of the environment individuals were subjectively representing while solving a particular maze. While other markers of attention and awareness (for instance, behavioural or neurophysiological variables) could also be used, here we focused on direct awareness reports in order to relate our findings both to those of Ho and colleagues and to the subjective awareness reports used in consciousness science (e.g. the Perceptual Awareness Scale (Barnett et al., 2024; Overgaard & Sandberg, 2021; Ramsøy & Overgaard, 2004; Samaha et al., 2015)). Participants were instructed to maintain central fixation while planning (see dataset dSC 1), in line with previous empirical work using this task (Ho et al., 2022).

We first reanalyzed the data presented by Ho and colleagues (2022)(Ho et al., 2022) to examine the role of spatial attention in building task representations (datasets Ho 1 and 2). In a new experiment (dataset dSC 1), we designed novel mazes to test the effects of lateralization of attention in enabling efficient planning (see Methods & Table S1). In addition, we recorded the eye movements of participants during planning to ensure that any attentional effects observed were driven only by covert shifts in attention (see Methods).

We retained trials in which participants successfully navigated each maze (see Methods). Note that the successful navigation of the maze stimulus and the construal process represent two distinct processes. For example, we can imagine a trial in which a participant represents every single obstacle, whether relevant or irrelevant. We would predict that on this trial, the participant could successfully navigate the maze, yet their construal process would be suboptimal according to the normative sVGC model.

Our focus in the present study was to examine attentional effects on participants' perception of the maze stimulus. We did not quantify how individuals deploy their attention in the phase in which they were navigating through the maze.

## A spotlight of attention influences task representations

We hypothesized that spatial attention would control which items are included in a task representation(He et al., 1996; Liu et al., 2009; Whitney & Levi, 2011). Specifically, we hypothesised that participants would deviate from the predictions of the VGC model and become distracted by task-irrelevant obstacles when they are presented in spatial proximity to task-relevant obstacles. Note that the task-relevance of obstacles is related to the maze's organization and computational model, and is not related to participants' subjective reports. To evaluate these predictions, we first computed the distance between a probe obstacle and every other obstacle in the maze. Second, we ranked the obstacles from the closest to the furthest from the probed item. Using the ranked obstacles, we trained a linear regression model to predict participants' awareness of the probed obstacle (in green) from their awareness of the remaining obstacles (in grey; Figure 1b).

Critically, we observed a significant effect of spatial context on task representations – an effect which is not predicted by the normative VGC model. Participants' awareness of a particular obstacle was positively predicted by the awareness of its close neighbours ($\beta_1$ = 0.26, SE = 0.01, 95% CI [0.25, 0.28]; $\beta_2$ = 0.29, SE = 0.01, 95% CI [0.27, 0.30]), whereas awareness of its furthest neighbours negatively predicted participant reports ($\beta_5$ = -0.13, SE = 0.01, 95% CI [-0.15, -0.12]; $\beta_6$ = -0.13, SE = 0.01, 95% CI [-0.15, -0.12]; see Table S2). In other words, the spatial context of an obstacle predicted whether it would be included in a simplified task representation – akin to a diffuse attentional spotlight which filters which aspects of the maze are available for planning. This effect remained significant for both task-relevant and task-irrelevant obstacles, and after controlling for the predictions of the VGC model (Figure S7 & Table S3, respectively). We observed the same effect in a separate experiment where participants planned their route upfront before navigating the mazes (i.e., dataset Ho 2, see Figure S8 and Table S4 & S5). Finally, we replicated this pattern of results in our in-person experiment: closest neighbours positively predicted the awareness of an obstacle ($\beta_1$ = 0.19, SE = 0.007, 95% CI [0.18, 0.21]), whereas furthest neighbours negatively predicted participants' reports ($\beta_3$ = -0.10, SE = 0.01, 95% CI [-0.11, -0.08]; $\beta_4$ = -0.26, SE = 0.007, 95% CI [-0.27, -0.25]; $\beta_5$ = -0.29, SE = 0.007, 95% CI [-0.30, -0.27]; see Table S6 & S7 and Figure S9).

Next, we explored whether the influence of neighbouring obstacles on task representations varied across individuals. To do so, we fit the regression model described above to quantify each participant's attentional spillover, and quantified the linear slope of the resulting beta coefficients. Negative slopes indicate a significant effect of attentional spillover on task representation. The influence of attention varied considerably across participants: while on average, participants' task representations were influenced by attention (mean effect = -0.08; s.d. = 0.04), a subset of participants showed minimal influence of attention on their task representation (i.e., flat slopes; Figure 1c).

We hypothesized that participants with the largest attention effects (i.e., most negative slopes) would also show sparser task representation (i.e., a "spotlight of attention" which is focused only on a subset of obstacles). To test this, we computed the sparsity of participants' task representations by estimating the variance of their awareness reports, with higher variance indexing those participants who report being very aware of some obstacles and unaware of others. In line with our hypothesis, we observed that participants who were most influenced by neighbouring obstacles also showed sparser task representations (dataset Ho 1: $\rho = -0.35$, $p < 0.001$; dataset Ho 2: $\rho = -0.49$, $p < 0.001$; dataset dSC 1: $\rho = -0.51$, $p < 0.01$; see Figure S10). To address concerns of overfitting, we tested whether the spatial attention effects observed in a lateralized set of mazes generalized to task representations of non-lateralized mazes and vice versa (dataset dSC 1). We observed that inter-individual differences in spatial attention effects in one condition predicted the sparsity of task representations in the other ($\rho = -0.48$, $p < 0.01$; $\rho = -0.42$, $p < 0.05$).

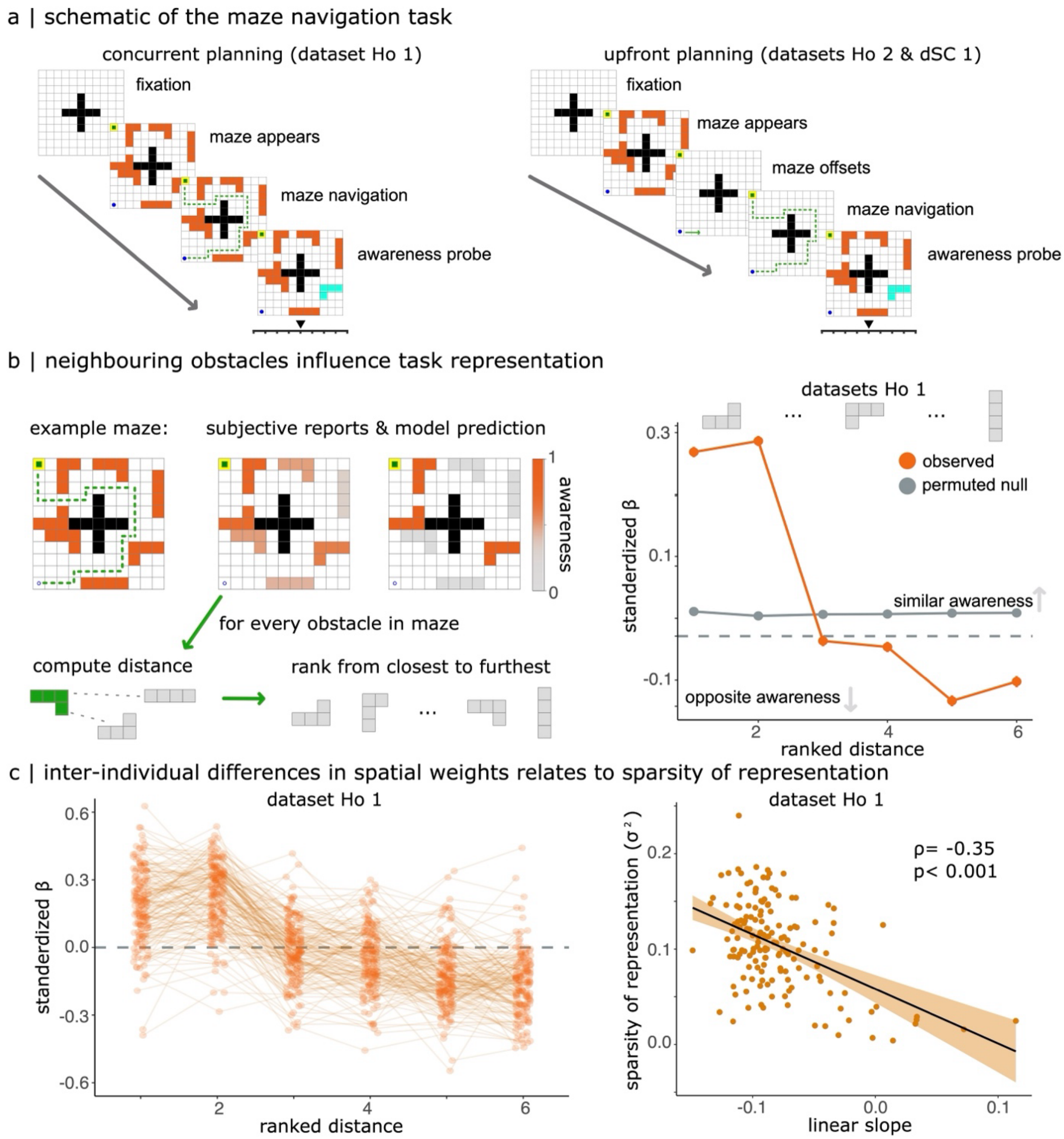

Fig. 1. Spatial attention shapes task representations.

(a) Schematic of the maze navigation task. Participants fixated at the start of each trial, after which a maze was presented, which they were asked to navigate. Maze stimuli either remained on the screen during navigation (left panel; *concurrent planning experiments*) or were removed before navigation (right panel; *upfront planning experiments*). Once participants finished navigating the maze, they were asked to report their awareness of every obstacle presented on a given trial in a random order.

(b) Left panel: schematic of the analysis pipeline. An example maze is shown where seven obstacles (plotted in orange) are presented on every trial according to pre-defined mazes. Participants report their awareness of every obstacle at the end of each trial (middle maze). The VGC model predicts which obstacles in a maze will likely be included in participants' task representation (right maze). We use participants' awareness reports to test the influence of neighbouring obstacles on the probe obstacle (presented in green). We compute the influence of neighbouring obstacles (in grey) on participants' awareness of the probed obstacle (in green). Right panel: Results of the ranked regression model for dataset Ho 1. We observed that obstacles closest to the probed item (rank 1 & 2) positively impact awareness reports. In contrast, obstacles furthest from the probed item negatively impact awareness reports (rank 5 & 6).
(c) Left panel: The effect of neighbouring obstacles on task representations varied across participants (each represented by a point). Right Panel: Inter-individual differences in the attentional effects correlate with the sparsity of participants' representations. Participants who showed the greatest influence of neighbouring obstacles (more negative slopes), showed the simplest representations (greatest variance in awareness reports).

### Attentional limits constrain the optimality of task representations

Prior psychological research indicates that attention can be efficiently allocated to a "hemifield" of visual space – with information being preferentially processed when presented in the attended hemifield(Eriksen & St. James, 1986; M. I. Posner, 1980; *The Principles of Psychology, Vol I.*, n.d.). Building on this work, we hypothesized that participants would select task-relevant information with greater ease – constructing task representations more closely aligned with the VGC model – when task-relevant information is spatially confined to a visual hemifield (i.e., presented unilaterally).

To test this hypothesis, we derived a measure of task-relevant lateralization inspired by the attention literature (Ghafari et al., 2024; Keefe & Störmer, 2021; Vollebregt et al., 2015) (Figure 2a). Specifically, we separated maze stimuli across the vertical meridian and computed the ratio of task-relevant information presented on the left versus right side derived from the sVGC model. For example, the maze shown in Figure 2a has twice the amount of task-relevant information presented in the left hemifield than in the right (lat. Index= 1/3). A lateralization index of 0.0 indicates that both hemifields contain equal amounts of task-relevant information (i.e., non-lateralized). The lateralization index was computed using the continuous VGC predictions for each obstacle (see Methods). We used this task-relevance lateralization index as a moderator in a hierarchical linear regression model to test whether participants' awareness reports were better predicted by the original VGC model in mazes showing the greatest lateralization of task-relevant information. In addition, we monitored participants' eye movements in dataset dSC 1 to ensure that attention shifts would be covert as opposed to overt—a distinction which could not be determined in the online samples of datasets Ho 1 and 2.

In line with our hypothesis, we observed a significant moderation effect whereby the greater the lateralization of task-relevant information across the vertical meridian, the better the original VGC model was at predicting participants' awareness reports ($\beta_{interaction}$ = 0.01, SE = $2.65*10^{-3}$, 95% CI [0.01, 0.02], $p_{FDR} < 0.001$; Figure 2c left panel & Table S8). We replicated these findings with the data collected in dataset Ho 2(Ho et al., 2022) ($p_{FDR} < 0.01$; see Table S9). These results indicate that participants' task representations are more closely aligned with the ideal observer (i.e., the original-VGC model) when task-relevant information is presented unilaterally.

In our new dataset (dSC 1), we designed novel maze stimuli to validate these lateralised effects of attention while addressing some limitations of previous experiments (see Methods). We again observed that lateralization of task-relevant information impacted participants' awareness reports. Participants were less aware of task-irrelevant stimuli on trials where the lateralization of task-relevant information was larger (Figure 2b) and we replicated the moderation effect of information lateralization on the extent to which the original VGC model captured participants' awareness reports ($\beta_{interaction}$ = 0.01, SE = $2.65*10^{-3}$, 95% CI [0.01, 0.02], $p < 0.001$; Figure 2c & Table S10). This effect did not vary significantly as a function of the specific hemifield (i.e., left vs right) in which task-relevant information was presented (β= 0.01, SE= 0.02, 95% CI [-0.03, 0.04], p = 0.738; ΔBIC = 58.30 in favour of the null effect; see table S14).

We note that for three maze stimuli whose model predictions were lateralized there was nevertheless a poor fit to the sVGC model (see Figure 2c, right panel). These outliers correspond to maze stimuli where participants, on average, lateralized their attention to the incorrect hemifield (i.e., the opposite hemifield to that predicted by the sVGC model).

In contrast with our observations of consistent and strong attentional effects relative to the vertical meridian, effects relative to the horizontal meridian (superior vs. inferior) were inconsistent across experiments. Specifically, we observed a significant moderation effect in dataset Ho 2 ($\beta_{interaction}$ = 0.01, SE = 2.85*$10^{-3}$, 95% CI [0.00, 0.01], $p_{FDR}$< 0.05; see Table S9), but not in dataset Ho 1, and the moderation effect was negative rather than positive in dataset dSC 1 ($\beta_{interaction}$ = -0.01, SE = 2.22*$10^{-3}$, 95% CI [-0.01, 0.00], $p$< 0.05). The effect in the Ho 2, but not the dSC1, dataset became insignificant after accounting for nuisance covariates (see Table S15 & S16).

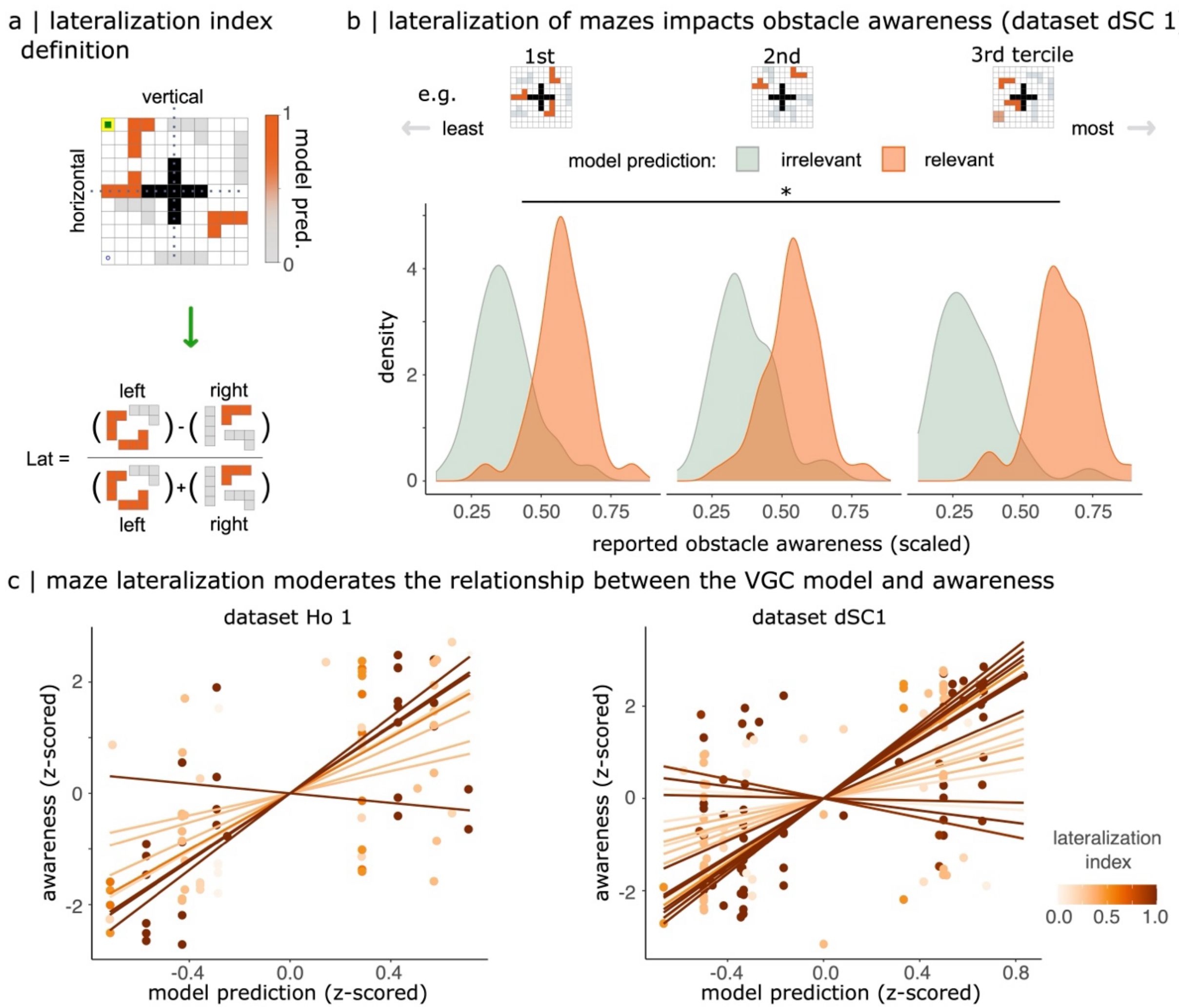

Fig. 2. Lateralization of task-relevant information affects task representations.

(a) For each maze, we computed a vertical meridian lateralization index. This index reflects whether task-relevant information is lateralized to a hemifield. In the example plotted, there is more task-relevant information presented on the left than on the right of the maze, therefore this would correspond to a moderate level of vertical meridian (i.e., left vs right) lateralization. We similarly computed an attention index for the horizontal meridian (i.e., above vs below).
(b) Density plots of the reported awareness of obstacles on the basis of whether the value-guided construal (VGC) model predicted them to be task-relevant (≥0.5; in orange) or task-irrelevant (< 0.5; in grey). Note sVGC model predictions for each obstacle were binarized for visualization purposes only. Participants were more likely to be aware of obstacles predicted as task-relevant. We split maze stimuli based into terciles based on the degree to which task-relevant information was presented preferentially to one hemifield (x-axis). The leftmost plots are mazes where task-relevant information is presented on both hemifields. In contrast, the rightmost plot depicts mazes with the largest lateralization. We observed that the awareness reports of participants become increasingly aligned to the VGC model's predictions as lateralization increases.
(c) Scatter plot of the effect of maze lateralization on the relationship between the value-guided model and participants' awareness of obstacles. We observed a significant vertical meridian lateralization effect whereby participants' awareness reports were more strongly aligned with the VGC model's predictions when task-relevant information was presented unilaterally in all datasets. Each point represents an obstacle in a maze, and each line represents the model fit for that specific maze stimulus.

### Inter-individual variation in lateralization of attention

Next, we investigated participants' tendency to pay attention to obstacles within a single hemifield (left vs right) regardless of the sVGC model predictions. To do so, we computed an awareness lateralization index (ALI) based on participants' self-reported awareness reports of obstacles on each trial (Figure 3a). Large positive values indicate that participants were preferentially aware of the right hemifield, whereas negative values indicate preferential awareness of the left hemifield. Values close to zero indicate that participants paid attention to both hemifields equally (see Methods for details). We observed that participants' tendency to lateralize their awareness varied greatly across the Ho datasets 1 and 2 (Figure 3b); some participants preferentially paid attention to a single hemifield, regardless of whether the sVGC model predictions were lateralized. For the dSC1 dataset, we observed that on some trials, participants significantly lateralized their awareness (|ALI| > 0.5; Figure 3c) even though the sVGC model predictions were non-lateralized. These findings suggest that participants' tendency to pay attention to a single hemifield may represent an observable inter-individual difference in how they allocate their awareness to form task construals.

To further explore these inter-individual differences, we tested whether participants' tendencies to lateralize their attention to a single hemifield was consistent across trials and maze stimuli. We observed that participants' tendency to lateralize their attention to a single hemifield was similar for left and right lateralized maze stimuli (Spearman $\rho$= 0.72, Figure 3d). This suggests that participants who preferentially attended to a single hemifield did so regardless of which hemifield they should attend to. More consequentially, the tendency for participants to lateralize their awareness on maze stimuli whose model predictions were also lateralized linearly correlated with participants' tendency to lateralize their attention on non-lateralized maze stimuli (Spearman $\rho$= 0.88, Figure 3d). Taken together, these findings emphasize that some individuals tend to preferentially attend to a single hemifield when planning. This tendency, importantly, represents an inter-individual difference in how participants allocate their attention across various maze types.

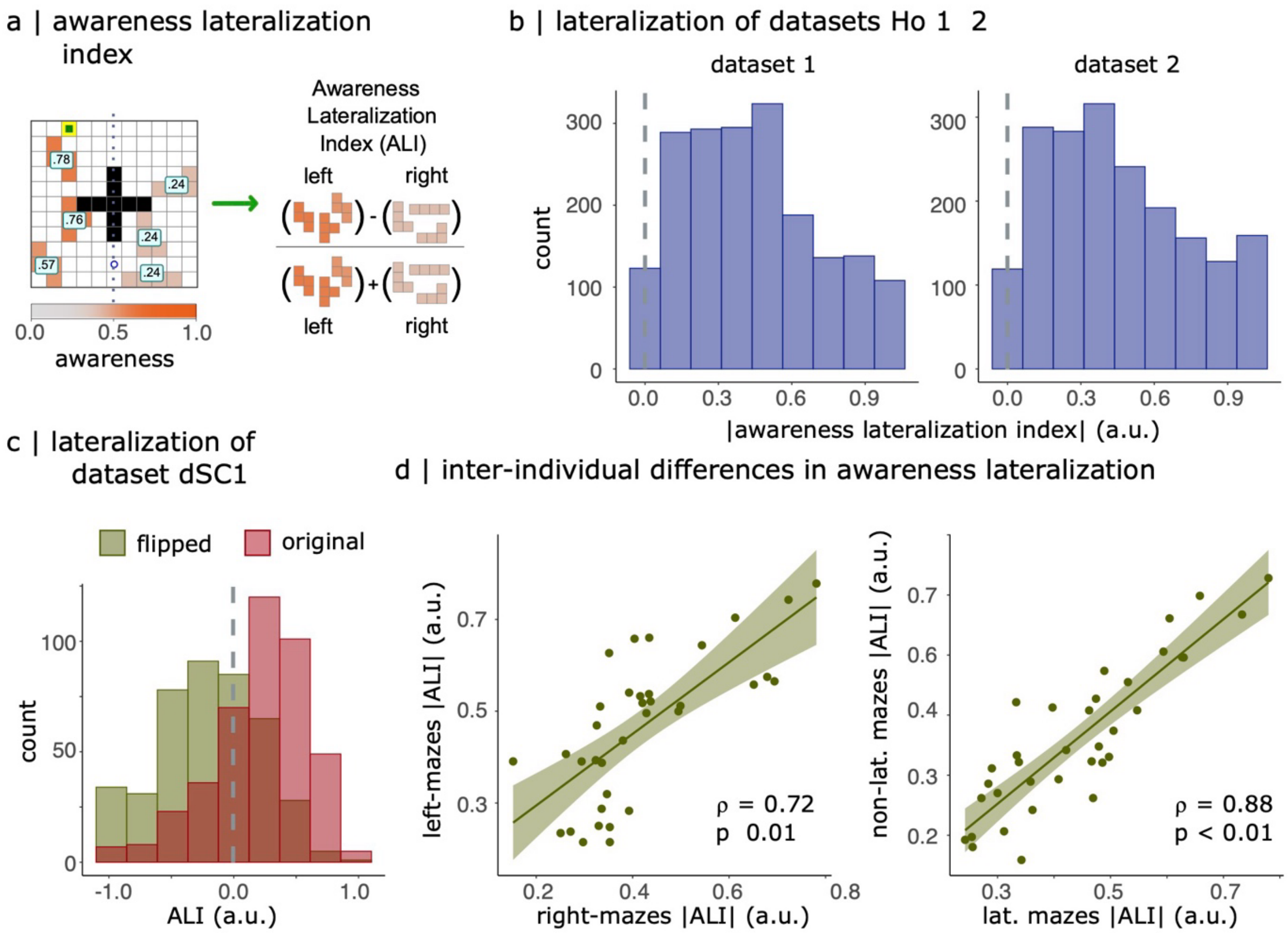

Fig. 3. Inter-individual variation in lateralization of awareness.

(a) For each maze and participant, we computed an awareness lateralization index (ALI). This index reflects the degree to which participants tended to pay attention to obstacles in a single hemifield. In the example plotted, the participant preferentially paid attention to the obstacles presented to the left hemifield regardless of whether they were task-relevant or task-irrelevant. Note that this lateralization index is based on participants' self-reports, unlike the lateralization index presented in Figure 2 which is based on the sVGC model predictions.

(b) Histogram of the ALI of participants across the Ho 1 & 2 datasets. In these experiments, some participants showed substantial lateralization of awareness (ALI > 0.5), despite the maze stimuli for these experiments being—on average—non-lateralized in their sVGC model predictions.

(c) Histogram of the ALI of participants for maze stimuli with non-lateralized VGC model predictions. We plot ALI values separately for the original non-lateralized mazes, and the left-right reversed (flipped) mazes separately. Participants on average did not lateralize their awareness. We note, however, that on some trials participants' awareness reports were strongly lateralized, which contrasts with the sVGC model predictions.

(d) Scatter plot of participants' tendency to lateralize their attention to either hemifield (i.e., absolute value of ALI). We plot this for mazes with left and right lateralized model predictions (left panel) and for mazes with non-lateralized and lateralized model predictions (right panel). The large linear relationships indicate that participants' tendency to lateralize their awareness is a stable inter-individual difference.

## Attentional spotlight model of task representations

Taken together, our results corroborate a critical role for visuospatial attention in constructing task representations. Notably, these filtering effects of attention on value-guided construal are not currently part of the original VGC framework proposed by Ho and colleagues. In what follows we explicitly incorporate the influence of a spotlight of attention into the original VGC model to formulate the spotlight-VGC model(Eriksen & St. James, 1986; M. I. Posner, 1980).

To achieve this, we computed the predictions of the existing VGC model for each obstacle's task relevance in a given maze, and averaged these predictions within an attentional spotlight of 3 squares (Figure 4a & S8, see Methods for details). This process yielded novel model predictions, whereby some obstacles which were once predicted as task-irrelevant by the normative sVGC model are now predicted as task-relevant by the attentional spotlight model. We depict the effects of this spatial spotlight in Figure 4a: task-irrelevant stimuli (plotted in grey; see middle left obstacle) neighbouring task-relevant obstacles (plotted in orange) become more task-relevant, whereas task-relevant information becomes less relevant when surrounded by task-irrelevant information (see bottom right orange obstacle). This deviation in model predictions from the normative sVGC model was used to predict participants' awareness reports. We hypothesized that this spotlight-VGC model would predict participants' reports better than the original VGC model, which does not account for spatial attention.

In line with this hypothesis, we observed that the spotlight-VGC model predicted participants' awareness reports better than the original VGC model in all three datasets (dataset Ho 1: ΔBIC= 84.63; Ho 2: ΔBIC= 203.43; dSC 1: ΔBIC= 70.72; see Figure 4b right panel). For dataset dSC 1, we observed a significant improvement in model fit for non-lateralized maze stimuli (ΔBIC= 161.93) but failed to find any improvement when maze stimuli were lateralized (ΔBIC= -42.02; see Figure 4c). These findings dovetail with the previously discussed moderation effects, and suggest that the spotlight-VGC model is particularly useful in improving our ability to explain human behaviour in situations when attentional filtering is more complex.

To further explore inter-individual differences in task construal, we tested whether adjusting the attentional spotlight width to each participant's awareness reports improved the predictions of the attentional spotlight model. To do so, we first determined the width attentional spotlight of each individual in the dSC1 dataset based on lateralized maze stimuli. We then generated person-specific attentional spotlight model predictions for the non-lateralized maze stimuli to avoid overfitting the data (Figure S11). We note that 7 participants had either flat attentional slopes or negative beta coefficients, which prevented the selection of an appropriate attentional spotlight width (see Methods for details). We observed a significant improvement in model fit for the person-specific attentional spotlight model relative to both the group-level attentional spotlight model ($\Delta$BIC= -1487.39) and the normative sVGC model ($\Delta$BIC= -1655.29). While the limited trial numbers per participant in our current dataset warrants caution in interpreting these findings, these findings do encourage further research on inter-individual differences in attentional deployment during planning.

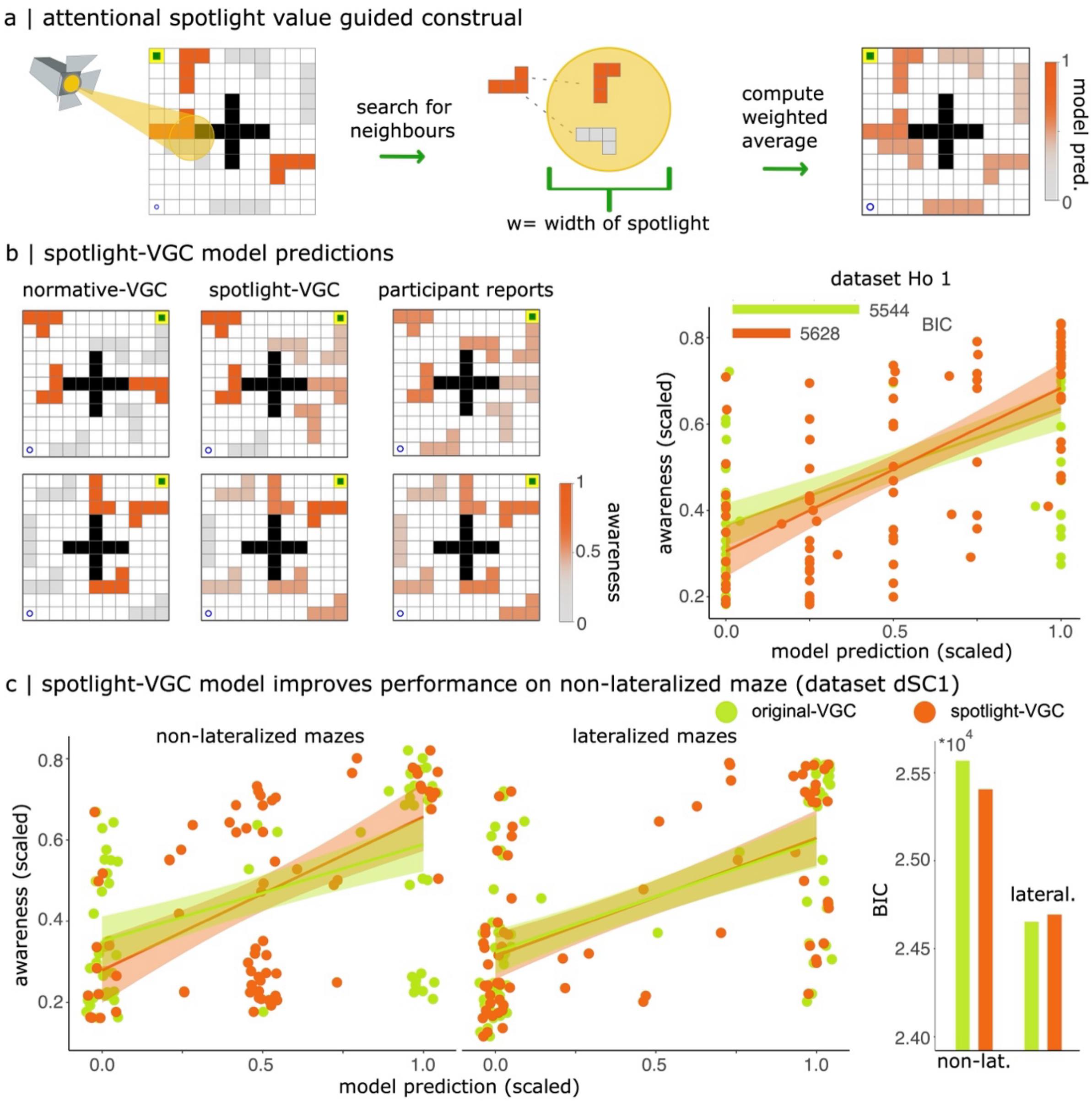

Fig. 4. A VGC model augmented with an attentional spotlight model predicts participants' task representations.

(a) Schematic of the attentional spotlight model. Inspired by the spotlight of attention analogy, we recompute an obstacle's probability of being included in a task representation as the weighted average of its neighbours. We first search for all neighbours of $obstacle_i$ that are w squares away. We then compute $P(Obstacle_i)$ as the weighted average of $obstacle_i$ and its neighbours. This generates more graded model predictions (far right panel).

(b) Left panel: Each row represents a different example maze stimulus. The left column depicts the original VGC model prediction $P(Obstacle_i)$ for every obstacle in the example maze. The middle column shows the attentional-spotlight model prediction for every obstacle. Obstacles that were considered task-relevant (deep orange) in the original model become less important when surrounded by task-irrelevant information (grey obstacles). The right column shows the participants' average awareness of each obstacle in the example mazes. Right panel: Scatter plot of the linear relationship between participants' awareness reports of obstacles and model predictions (original-VGC in green and the spotlight-VGC model in orange) for dataset Ho 1. The latter fits participants' reports better than the original VGC model.
(c) Scatter plot of the linear relationship between participants' awareness reports of obstacles and model predictions (original VGC in green and the spotlight-VGC model in orange) for dataset dSC 1 separately for non-lateralized (left panel) and lateralized mazes (right panel). Although both models fit participants' awareness reports better for lateralized mazes, the advantage of the spotlight model over the original model (better model fit / lower BIC) was observed only in non-lateralized mazes.

## Maze navigation performance

The previous analyses focused on participants' task representations during planning. We next sought to explore links between participants' task representations and maze navigation performance. Participants performed the maze navigation task near-ceiling: they solved 95% of maze stimuli in under 20 seconds, with minimal deviation from the optimal path (i.e., 9 moves or fewer). Notwithstanding this limited variance in task performance, we explored whether participants' task construals may have impacted their navigation speed. To do so, we first regressed out the effects of the sVGC model from participants' awareness reports and used the mean squared residuals for each trial to predict response times (see Methods for details). Surprisingly, we observed a negative relationship between mean squared residual variance and response times ($\beta = -0.31$, SE = 0.05, 95% CI [-0.41, -0.21], $p < 0.001$), indicating that participants were faster on trials where the sVGC model explained less variance in their awareness reports. In other words, trials in which participants deviated *more* from the sVGC model predictions were solved faster. We note that one reason for this may be the strong influence of the lateralisation effect on navigation performance (see paragraph below), which itself is not part of the sVGC model prediction.

We then explored whether participant performance differed between lateralised and non-lateralised mazes. Here, we reasoned that the initial phase of lateralised attentional selection would lead to lateralised mazes being easier to navigate than non-lateralised ones. Consistent with this hypothesis, participants were faster ($\beta = -0.04$, SE = $5.91*10^{-}$

$^{3}$, 95% CI [-0.06, -0.03], $p< 0.001$) and followed the optimal path more closely (β = -0.59, SE = 0.09, 95% CI [-0.78, -0.40], $p< 0.001$) when maze stimuli were more lateralized.

## Sensitivity analyses

We conducted a series of control analyses to verify the robustness of our experimental results. First, we verified that the spatial proximity effect (Figure 1b) was not driven solely by the spatial smoothness of participants' awareness reports by conducting null permutation tests (grey line, Figure 1b). For each maze stimulus, we permuted the rank of the neighbouring obstacles. We then fit the same linear model to assess the effect of spatial context on task representations. This procedure was repeated 1000 times to generate a null distribution of beta coefficients. The resulting null distribution showed no discernible effect of spatial context. Second, we used null permutation tests to verify that the improved fit of the spotlight-VGC model was not driven by greater spatial smoothness of the model predictions (see Supplemental Figure S12). Third, we assessed whether nuisance covariates could explain the moderation effects we observed. Specifically, we added the distance from the goal, starting location, center walls, and fixation as nuisance covariates in our hierarchical regression models. Maze lateralization remained a significant moderator of the relationship between the original VGC model and participants' awareness reports after controlling for these covariates (see Table S11-13). This was not the case, however, for lateralization effects along the horizontal meridian (see Table S15-16).

Fourth, we sought to verify that the lateralization effects we observed were not driven by a change in eye movement patterns. For dataset dSC 1, we continuously tracked the position of participants' gaze. We explicitly instructed participants to maintain central fixation while planning (see Methods for details) and removed the obstacles from the screen after 6 seconds. This allowed us to verify that greater awareness of obstacles was not driven by longer fixation times. We confirmed that participants maintained central fixation on both lateralized and non-lateralized maze stimuli in most trials (see Figure S13). Excluding trials where participants exhibited excessive eye movements during planning, we continued to observe qualitatively similar lateralization effects (see Figure S15 and Table S17).

Finally, we examined the convergent validity of participants' awareness reports by reanalyzing the memory recall data reported in Ho and colleagues' experiment (Ho et al., 2022). We reasoned that participants should demonstrate similar task representations regardless of the measure used to probe the construal. In line with this prediction, we observed that the obstacle awareness reports and memory/hover measures were strikingly correlated within three independent samples of participants (Spearman $\rho = 0.86$ between memory accuracy and awareness; $\rho = 0.86$ between confidence in memory and awareness; $\rho = 0.76$ between the probability of hovering over the obstacle and awareness; $\rho = 0.65$ between the duration of the mouse hovering and awareness; see Tables S18 and S19).

## Discussion

Searching for a solution in a complex multi-step task is challenging. Recent computational work suggests that humans overcome this challenge by constructing simplified perceptual representations of their environment. In the present study, we reveal a role for visuospatial attention in constructing these simplified perceptual representations.

### Participants' task-representations are informed by visuospatial attention

We provide several lines of evidence for the critical role of visuospatial attention in constructing task representations. First, we observed a significant effect of the spatial context in which information is presented. Participants were less likely to incorporate task-relevant information into their construal when it was surrounded by task-irrelevant information. These effects mirror perceptual crowding effects(Liu et al., 2009; Whitney & Levi, 2011) which reveal that attention spills over to distractors presented alongside task-relevant stimuli when presented in close spatial proximity. Second, we observed that participants incorporated task-relevant information into their task representations more frequently when relevant obstacles were grouped together within the same hemifield (Figure 2). Participants' task representations in such settings were more closely aligned with an ideal observer model – suggesting that the natural contours of visuospatial attention interact with the capacity of observers to form efficient task representations.

### Incorporating attention into a model of value-guided construal

The VGC model articulates how an ideal decision-maker should represent an environment while balancing complexity and utility(Ho et al., 2022). We developed an extension of this model that accounts for the effects of spatial attention on planning. Our model, inspired by the analogy of a spotlight of attention(Carrasco, 2011; M. I. Posner, 1980), provides a better fit to participants' awareness reports than the original VGC model (Figure 4). This improved model fit was most evident for mazes where task-relevant information was presented to both hemifields (Figure 3c), suggesting the augmented model is helpful in explaining behaviour in contexts where attentional selection is more complex. These deviations from the original VGC model, therefore provide a useful benchmark to compare human performance and offer insights into natural constraints on human cognition. For instance, we demonstrate that spatial context biases whether information is to be included or excluded from a representation of the environment. These effects may reflect inductive biases in humans who have learned and evolved to select information from real-world environments where obstacles tend to be grouped together in visual scenes(Kaiser et al., 2019; Peelen & Kastner, 2014). However, it is plausible that these inductive biases on value-guided construal may themselves be learnt, and vary according to other environmental demands and contexts which impose systematic regularities on useful task representations (e.g., attending preferentially to intersections when planning on the Tube). Future research can explore the flexibility of participants' task representations across environmental contexts, and ask how these inductive biases are acquired.

### Inter-individual differences in attention

We also observed considerable inter-individual differences in attentional effects across participants (Figure 1c). While some participants were strongly influenced by the spatial context of neighbouring stimuli, others showed more limited evidence for an attentional effect (Figure 1b). Inter-individual differences in attention predicted the sparsity of participants' simplified representations: participants with larger attention effects exhibited sparser representations. Moreover, these inter-individual differences in effects of spatial proximity could be incorporated into the attentional spotlight model by varying the width of the spotlight, resulting in better model predictions.

Beyond these spatial proximity effects, we also observed that participants varied in their tendency to lateralize their attention to a single hemifield (Figure 3). This tendency was observed across all three datasets, including on maze stimuli whose value-guided model predictions were not lateralized. This suggests that although a strategy of allocating attention is sub-optimal for these maze stimuli, some individuals preferentially attend to a single hemifield in a heuristic-like fashion. This tendency to attend to a single hemifield was a robust inter-individual difference across maze stimuli (Figure 3d), and dovetails with individual-level variation in spatial proximity effects. Taken together, these findings offer novel insights into how people vary in the ways they allocate spatial attention to solve complex problems. Future research could explore how these individual differences constrain performance on other tasks that require planning and search in high-dimensional spaces.

### Mental representations and task performance

We observed that participants were faster and deviated less from the optimal path on maze stimuli that were lateralized. This effect is not predicted by the original sVGC model but dovetails with the interpretation that early visuospatial attention operates as an inductive bias to guide the formation of simplified task representations. Surprisingly, we also observed that participants were faster to navigate mazes on trials where their simplified task representation deviated from the sVGC model prediction. We interpret this seemingly contradictory finding in the following way: there are several factors beyond the sVGC model – including, for instance, maze lateralisation – that predict both construal and performance on the maze navigation task. Further work is needed to understand how inductive biases such as lateralisation shape both construal and performance, and the real-world benefits that such strategies might afford for naturalistic stimuli.

## Planning and Consciousness

Our experiments investigate the connection between planning and participants' reports of their awareness of features of the task environment. The results may therefore be relevant to understanding the functions of conscious experience. While an intimate connection between attention and consciousness is widely recognized(Cohen et al., 2012; Koch & Tsuchiya, 2007; V. A. F. Lamme, 2003; Tsuchiya & Koch, 2014), there is less work explicitly considering the connection between planning and consciousness(Fleming & Michel, n.d.; MacIver & Finlay, 2022). However, there are several reasons why the kind of planning at work in our experiments is likely to require the task to be represented consciously.

As we mentioned at the outset, simplified representations reduce the computational burden of planning in a branching multi-step task space. The same consideration suggests that planning should be based on conscious, rather than unconscious, representations(Shea & Frith, 2016). Initial stages of perceptual processing can carry information about a range of different and incompatible possibilities at once, for example, a probability distribution across a range of possible orientations of a line. The probabilistic representation attaches some probability (or probability density) to many different possibilities. There is, of course, a certain burden in integrating and weighing probabilistic information of this kind, for which the brain is thought to deploy various solutions (e.g. approximate Bayesian inference). These initial stages of perceptual inference are typically thought to be unconscious. However, forward planning from multiple possibilities in a branching task space rapidly becomes intractable as the combinatorial possibilities explode. Consciousness, by contrast, provides a much sharper representation of the current state, from which planning can proceed forward(Block, 2018; Stocker & Simoncelli, 2007).

Given the computational cost of running through and comparing many potential multi-step action sequences, it makes sense to base that process on a reliable estimate of the current world state. While it is doubtless useful to produce some kinds of unlearned and habitual action very rapidly at the first hint of information, for example of the presence of a predator, with multi-step forward planning it makes sense to integrate information from more sensory modalities and across a longer timescale before then committing to using a representation as the basis for planning. This again suggests that conscious representations, which are known to integrate information across modalities and time(Bekinschtein et al., 2009; Deroy et al., 2016; Herzog et al., 2020; Mudrik et al., 2014; Strauss et al., 2015), are perfectly suited to the functional needs of this kind of planning task. Furthermore, planning depends on both facts and values. Potential actions are assessed based on the expected value of outcomes. The role of value was captured, in our studies, by an extension of the VGC (value-guided construal) model. Consciousness is thought to facilitate the integration of different sources of value(Braver et al., 2014; Dickinson & Balleine, 2010; Dung, 2022).

The task and associated computational model thus offer a flexible tool for characterising the computations by which conscious representations influence decisions and actions. Future work could tell us more about the way bottom-up attention-driven inputs and task-based value jointly influence what information reaches conscious experience. This provides a novel ecologically-valid probe of the connections between attention, consciousness, and decision-making that does not require the explicit labelling of task-relevant stimuli. Neural (e.g. M/EEG) data collected while participants plan could help understand the timescale and computational steps that lead to the formation of a conscious task representation. Modifications of the paradigm would also be suited to exposing the way non-consciously-presented cues do and do not influence the way participants plan.

## Methodological Considerations and Future Directions

We close by reflecting on opportunities for further work in this area. First, an important next step is to explore the process by which task representations are formed, and how inductive biases might affect the process of task construal. The sVGC model is a normative model of the optimal task representation. Since it’s construction involves an exhaustive calculation over possible paths, it is not a plausible basis for a model of the psychological process by which participants actually construct task representations. More recently a process model of task construal has been proposed, the Just in Time model (JIT). The hypothesis of the JIT model is that participants’ task representations are built up over time by iteratively simulating possible paths through the maze, affording insight into the construal process (Chen et al., 2026). In future work, it would be of interest to ask whether the attentional effects we observe in our experiments could be meshed with a dynamic JIT account of construal. We speculate that visuospatial attention may operate as an early filter, limiting the space of potential construals based on coarse spatial features of the environment, constraining a dynamic selection of obstacles. Brain imaging techniques with high time resolution, such as M/EEG, may be able to shed further light on how task representations are formed as participants plan.

Second, in the current work we were unable to distinguish whether these attentional effects are driven by a fixed spotlight of attention, or whether attention operates akin to a zoom lens, shifting the ‘width’ of the focus of attention according to the task demands (Eriksen & St. James, 1986; Müller et al., 2003; Schad & Engbert, 2012). The latter view would be consistent with growth-cone models of attention in which the focus of attention expands and contracts in accordance with task demands, mirroring the various receptive field sizes in the visual hierarchy (Pooresmaeili et al., 2014; Pooresmaeili & Roelfsema, 2014). In partial support of this idea, we found significant inter-individual differences in the width of participants’ attentional spotlight (Figure S11). It is also possible that attention is deployed within or along parts of obstacles, rather than on entire obstacles. Future work using naturalistic measures of eye movements may be able to address these questions.

Third, while we observed clear lateralization effects along the vertical meridian (i.e., left vs right hemifield), effects along the horizontal meridian were less clear (i.e., above vs below; see Table S15-16). One potential explanation of this asymmetry is the retinotopic organization of the cortex, in which spatially adjacent stimuli can be retinotopically distant if presented on the opposite side of the vertical (but not horizontal) meridian, facilitating distractor inhibition. Importantly, while the visuospatial attention effects observed in the Ho 1 and 2 datasets are likely driven by both covert and overt shifts in attention, the findings presented in experiment 3 (i.e., dSC1 dataset) rule out the contribution of overt shifts in attention through the use of eye tracking (see Figure S13-14)(Carrasco, 2011; Pooresmaeili & Roelfsema, 2014).

Fourth, it will also be necessary to elaborate on how bottom-up and top-down aspects of attentional selection are combined to guide complex task representations and plans. Foundational questions remain unanswered, for instance: can multiple spatial locations be preferentially selected at once, i.e. are there multiple spotlights (Awh & Pashler, 2000; McMains & Somers, 2004; Pylyshyn & Storm, 1988; Shaw & Shaw, 1977)? There is also discourse on how spatial attention may move from one location to another: are the intervening visual regions between attended locations similarly selected (Dubois et al., 2009; Kr & Np, 1999; McMains & Somers, 2004, 2005)? Our findings tentatively suggest that individuals are able to attend to disparate spatial regions to form sparse task representations, yet there is substantial variability in how individuals orient their attention during the task. The present paradigm and computational modelling, in conjunction with carefully designed stimuli, may help resolve these outstanding questions.

Finally, our present study focused on studying mental representations for planning in the context of a navigation task. Whether these effects hold across other forms of planning, including planning over abstract spaces, remains to be demonstrated. An important next step to further our understanding of task representations would be to extend the current paradigm to other forms of planning and more naturalistic tasks, such as navigating immersive virtual reality (VR) environments, planning over cognitive rather than perceptual representations (e.g.. planning over an abstract space), or internally-guided planning based on working memory. In this spirit, recent work has applied the VGC model to a physical reasoning task in which participants were asked to predict the trajectory of a blue ball (Chen et al., 2026). Future work could also profitably examine the relevance of visuospatial attention for the navigation process itself in this task. While our present findings speak to how individuals perceive the maze while planning, it remains unclear how attention is deployed during navigation along a path, such as how object-based attention progressively spreads along trajectories in time and space (Pooresmaeili & Roelfsema, 2014; Wong & Scholl, 2024).

## Conclusions

Complex daily decisions require a decision-maker to arbitrate over countless potential multi-step actions and their outcomes, making searching for a solution difficult. We shed light on how this is achieved by clarifying the role of visuospatial attention in forming simplified perceptual representations to aid in planning. We build on previous work on the effect of task relevance and develop a computational model which explicitly incorporates the role of attention in value-guided construal. Our model bridges the literature on perception, attention and computational models of planning to provide a more complete computational account of human cognition. We believe the results of this paper can inform future research on a comprehensive theory of human cognition and inspire novel biologically-informed intelligent algorithms.

# Methods

Experimental Task: To test our hypotheses we relied on a previously established maze-navigation task where participants are asked to move a circle avatar from a starting location to a goal using the arrow keys(Ho et al., 2022). Each maze consisted of an 11x11 grid with blue obstacles (7 obstacles in datasets Ho 1 & 2, and 6 obstacles in dataset dSC 1), and black central walls arranged in the shape of a fixation cross. Each trial began with a fixation cross (center walls), after which participants were prompted to navigate to the goal. Experiments differed in terms of i) the mazes participants navigated, ii) whether the obstacles were presented before or during the execution of the plan, and iii) what the participant reported.

We reanalyzed the data of Ho and colleagues' experiments 1 and 2 for the present study. In experiment 1 (i.e., dataset Ho 1), participants were presented with the obstacles throughout the trial. At the end of each trial, participants were asked to rate "How aware of the highlighted obstacle were you at any point?" using a nine-point scale. In experiment 2 (i.e., dataset Ho 2), participants were similarly asked to rate their awareness of the various obstacles but were required to plan their solution before they began to solve the maze. See (Ho et al., 2022) for details concerning the experimental procedures.

We did not reanalyze the results of the fourth experiment by Ho and colleagues. In this experiment, participants were not presented with all the information (i.e., obstacles) at once to solve the maze. Instead, they discovered obstacles by hovering over them with a cursor.

To further test the effects of attention on task representations, we designed a novel set of maze stimuli. This consisted of 12 mazes with task-relevant obstacles lateralized to a hemifield (left or right) and 12 non-lateralized stimuli. Each maze consisted of six obstacles, three on each hemifield, none of which crossed the veridical meridian. This ensured that there were an equal number of obstacles for computing the lateralization index (see below). Maze stimuli of both sets were equated on several nuisance covariates (see Supplemental Table S1). Maze stimuli were vertically and horizontally reversed (i.e., left-right flipped) such that participants could not predict the location of the start or goal location. This resulted in four potential orientations of each maze across all 24 mazes, 96 trials in total.

The design of the in-person experiment (i.e., dataset dSC 1) closely followed the second experiment of Ho and colleagues(Ho et al., 2022). On every trial, participants were presented with a maze stimulus for 6 seconds, over which they were required to plan. The maze stimulus was offset, and participants were required to solve the maze after a one-second delay. On every trial, participants reported on their task representations using a nine-point awareness scale.

Participants: For datasets Ho 1 & 2, participants completed the task online on Prolific. In dataset Ho 1, 194 participants completed submissions, 161 of whom were included in the final sample after exclusions. In dataset Ho 2, 188 participants completed submissions, 162 of whom were included. Participants were excluded from analyses based on pre-registered exclusion criteria as detailed in (Ho et al., 2022). In short, participants were excluded if 20% or more of their trials were removed based on reaction times, or if they failed 2 out of 3 comprehension checks.

For dataset dSC 1, 35 participants (mean age = 23.14, SD = 5.35; 12 male) completed an in-person eye-tracking experiment (see Eye-tracking acquisition). None of the participants were excluded from the data analysis. We excluded trials where participants' reaction times were longer than 20 seconds, or where participants deviated more than nine moves from the optimal path (which reflected 3SD above the mean).

Ethics: All procedures were approved by the University College London ethics committee and adhered to the Declaration of Helsinki. Informed consent was obtained from each participant prior to each experiment.

VGC model: We fit the previously described VGC model to our maze stimuli(Ho et al., 2022). Briefly, this model computes the optimal simplified task representation such that it maximizes the utility of the representation while also minimizing the cognitive cost of keeping information in mind. This model assumes that a decision-maker combines a subset of cause-effect relationships to represent their environment in aid of planning. For every possible construal, the model computes the value of a representation:

$$\mathrm{VOR}(c) = U(\pi_c) - C(c).$$

where $U(\pi_c)$ is the utility of a construed plan $\pi_c$, and $C(c)$ represents the cost of keeping that information in mind.

A task representation is selected according to a SoftMax decision rule. We then compute a marginalized probability for each obstacle being included within a construal,

$$P(\text{Obstacle}_i) = \sum \|\varphi_{\text{Obstacle}i} \in c\| P(c),$$

where $\varphi_{\text{Obstacle}i}$ is the cause-effect relationship for $\text{obstacle}_i$, P(c) is the probability that the task representation is selected, and ||X|| is a statement which evaluates to 1 if X is true, and 0 when X is false. We use the values of $P(\text{Obstacle}_i)$ for every obstacle in a maze to predict participants' awareness reports. See (Ho et al., 2022) for a detailed explanation of the computational model.

We focused our analyses on the *static* version of the VGC model (i.e., sVGC), whereby task representations are assumed to remain stable across planning. Our choice was informed by the design of the experiment where participants were required to plan over all obstacles at once.

Spatial proximity effects: To examine how the spatial context of information influences participants' awareness reports, we ran a hierarchical linear regression model. First, for every obstacle in every maze, we rank-ordered all other obstacles based on spatial proximity. That is, the participant's awareness report of the closest item to $\text{obstacle}_i$ on the trial was used as a predictor of the participant's report of $\text{obstacle}_i$ in a hierarchical linear regression model. This yielded a regression model with 6 regression coefficients predicting participants' awareness reports based on spatial proximity:

$$\begin{aligned}\text{report}_{\text{obstacle i}} \\ = \beta_0 + \beta_1 * \text{report}_{\text{obstacle 1}} + \beta_2 * \text{report}_{\text{obstacle 2}} \ldots + \beta_{6*}\text{report}_{\text{obstacle 6}} \\ + (1\,|\text{MazeID}) + (1|\,\text{ParticipantID}) + \varepsilon\end{aligned}$$

where *(1| MazeID)* and *(1| ParticipantID)* are random intercepts of each maze and participant, respectively, and $\beta_1$ reflects the contribution of the closest obstacle to $\text{obstacle}_i$. We interpret any significant effects in this model as the influence of neighbouring stimuli on participants' representations. We also fit the above hierarchical linear regression model for each participant separately. We report these individual beta coefficients in Figure 1b.

To ensure that the above spatial proximity effects were not driven by the VGC model predictions, we regressed out the effects of VCG model predictions from participants' awareness reports, and used the residuals of the model as the dependent variable in a second regression where we similarly predicted the effects of neighbouring stimuli on representations.

We verified that these effects were not explained by the spatial smoothness of our data by conducting 1000 spatial null permutations. For every iteration, we permuted the mapping between each obstacle in a maze and their spatial location maintaining the number of neighbouring obstacles for every trial. We fit a hierarchical linear regression model using this permuted data and built a distribution of null beta coefficients to compare to our observed effects.

The sparsity of task representations: We sought to test the relationship between i) inter-individual differences in attention effects and ii) the sparsity of task representations. First, we estimated the magnitude of each person's attention effect by fitting a linear slope to the beta coefficients obtained (see Spatial proximity effects). A participant with a large negative slope, therefore, showed a larger effect of neighbouring obstacles on their representation. Second, we operationalized the sparsity of participants' simplified representation as the variance of their awareness reports. A participant with a sparse representation shows a high variance in their awareness of different obstacles in a given maze. Last, we tested the linear monotonic relationship between the sparsity of participants' representations and the attention effects using Spearman correlation.

Lateralization index: To test the effects of lateralization of task-relevant stimuli on participants' awareness reports, we developed a lateralized index of task-relevance inspired by the alpha-power attention literature(Ghafari et al., 2024; Keefe & Störmer, 2021; Vollebregt et al., 2015). We divided each maze into a right and left hemifield and computed the ratio of task-relevant obstacles on both sides:

$$Lat.index = \frac{\sum sVGC_{left} - \sum sVGC_{right}}{\sum sVGC_{left} + \sum sVGC_{right}}$$

where sVGC is the model's prediction of each obstacle task-relevance for that maze. Note obstacles only with a majority of its blocks within a single hemifield were considered (3 or more squares). This yielded an index of task-relevance lateralization for each maze stimulus. We repeated the above procedure to obtain an index of task-relevance lateralization for the horizontal meridian (superior vs inferior hemifield).

We tested whether the lateralization index moderated the relationship between the value-guided model predictions (sVGC) and participants' awareness reports using a hierarchical linear regression model.

$$report = \beta_0 + \beta_1 * sVGC + \beta_2 * Lat + \beta_3 * Lat * VGC + (1\,|Maze_{ID} + Participant_{ID})$$

where $\beta_3$ represents the interaction between the VGC model predictions and the lateralization index.

<u>Inter-individual differences in lateralization of awareness reports:</u> To examine inter individual differences in participants tendency to lateralize their attention to one henifiled (i.e., left vs right) while planning, we computed an awareness lateralization index (ALI) based on participants reports. For each trial, we compute the ratio between participant's awareness of obstacles presented on the right vs left hemifield for every trial:

$$Awareness\ Lat.index = \frac{\sum Awareness\ Report_{left} - \sum wareness\ Report_{right}}{\sum wareness\ Report_{left} + \sum wareness\ Report_{right}}$$

where negative values of ALI indicate that participants preferentially paid attention to obstacles presented on the left hemifield. We report the average absolute value of the ALI across participants for the Ho datasets 1 and 2. For the dSC1 dataset we computed the ALI for every participant for lateralized and non-lateralized maze stimuli. We compute the Spearman correlation between participants' tendency to lateralize their attention on non-lateralized mazes and lateralized mazes. A large correlation indicates that participants' tendency to lateralize their attention to a hemifield was consistent across both maze types.

Spotlight-VGC model: Inspired by previous literature comparing visuospatial attention to a spotlight that moves across the visual field, we developed an extension of the VGC model to account for the effects of attentional selection in forming task representations.

To do this, we recomputed the $P(\text{Obstacle}_i)$ as a weighted average of its neighbours. We computed the distance between every obstacle in the maze, and searched for obstacles with neighbours within 3 squares (Manhattan distance) away from $\text{obstacle}_i$. We fixed the 'width' of the attentional spotlight to a distance of 3 squares based on the observation that the two neighbouring obstacles positively predicted the awareness of a probe. We observed that the mean and median distance between neighbouring obstacles of the $2^{nd}$ rank (i.e., second closest) was 3 squares away for all mazes (Figure S15). We therefore opted to fix the value of the attention spotlight to 3 squares based on these observations. Future work utilizing this model should consider the statistics of their maze stimuli when deciding on the 'width' of the attentional spotlight. Neighbouring obstacles that fell within the attention spotlight were averaged as follows:

$$P(\text{Obstacle}_i) = \frac{P(\text{Obstacle}_i) + mean(\ P(\text{Obstacle}_n)\ )}{2}$$

where n is the number of obstacles that fall within the width of the attentional *spotlight* (i.e., neighbouring items). We repeat this procedure for all obstacles within each maze. If an obstacle did not have any neighbours, then the value of $P(\text{Obstacle}_i)$ remained identical to the value of original VGC model.

We used the outputs of the attention spotlight model in a hierarchical linear regression to predict participants' awareness reports, where we included participant and maze random intercepts:

$$report = \beta_0 + \beta_1 * At.Sp.Model + (1 \,|Maze_{ID} + Participant_{ID})$$

All linear regression models were fit with the *lmer* package in R.

Personalized spotlight model: To assess inter-individual differences in the width of the attentional spotlight, we aimed to test whether person-specific attentional spotlight model predictions outperformed a constant a model where we held the width of the attentional spotlight constant. To do so, we first used the beta coefficients obtained for each participant from the spatial proximity effects model. We then thresholded the betas and recorded for each participant at which rank their beta coefficient dropped below 0.05 (i.e., a small effect). We then used the median distance between obstacles (i.e., se Figure S15) of this rank as the width of the attentional spotlight for each participant. This resulted in a majority of participants with an attentional spotlight value of 3 and 4 (Figure S11). We note that 7 participants were excluded from these analyses: these reflect participants with flat spatial proximity slopes or a negative beta coefficient for the first rank. We then adjusted the attentional spotlight model predictions according to each participant's width, and used these model predictions in a hierarchical linear model to predict participants' awareness reports.

Null permutations: To ensure that the improved model fit of the attentional spotlight model was not driven by the spatial smoothness of our data, we conducted a series of control analyses where we permuted the model predictions within mazes.

To do so, we re-assigned the $P(\text{Obstacle}_i)$ of each obstacle in a given maze to a random item such that each obstacle was given a new model prediction. This permutation procedure maintains the distribution of $P(\text{Obstacle}_i)$ across obstacles for each maze, while randomizing the location of task-relevant information. We repeated this procedure for each maze separately. We then used these random model predictions to predict participants' reports using the same hierarchical linear model described in Spotlight-VGC model. We repeated this procedure 1000 times to generate a null distribution of beta coefficients. We compared the observed beta value for the spotlight-VGC model against this distribution. We note that averaging neighbouring obstacles before or after the permutation of the model predictions qualitatively yielded the same result.

Eye-tracking acquisition: For dataset dSC1, participants completed the computer task while their eye-position and pupil size were monitored using an EyeLink 1000 Plus eye tracker at 1000Hz (SR Research, Osgoode, ON). Participants were seated comfortably in a dimly lit room in front of a 24-inch monitor set to the resolution of 1,920 x 1,080 pixels at 60 Hz. Participants were positioned 60 centimetres away from the screen and rested their heads on a mount. Stimuli were presented on MATLAB 2019a using psychtoolbox (3.0.16), synchronized with the eye tracker. Before the start of the experiment, participants completed a standard 5-point calibration procedure. Drift correction was applied after every block.

Eye-tracking preprocessing & analysis: Eye-tracking data were preprocessed with the PuPL toolbox in MATLAB(Kinley & Levy, 2022). Impossible data points (i.e., gaze outside the screen's bounds) were removed, in addition to data 50ms before and 150 ms after eye blinks (identified by pupillometry noise(Hershman et al., 2018)). Segments of missing gaze position, up to 400ms long, were interpolated using cubic splines. We analyzed eye position data between -1000 ms and 6000 ms around the presentation of the maze, which corresponds to the planning window and the one second prior to planning. To verify that participants did not move their eyes more frequently during planning for lateralized mazes, we computed the standard deviation of eye position along the X-axis for each trial. We compared the fluctuations across lateralized and non-lateralized trials with a two-sample t-test. To verify the robustness of our behavioural effects, we identified and removed from further analysis trials where participants' eye position exceeded two squares away from fixation.

Navigation performance & mental representations: To examine how mental representations related to the navigation task we examined whether the lateralization of maze stimuli related to the time it took participants to navigate each maze (i.e., their response time). We ran a hierarchical linear regression model where we predicted the response time of each trial from the optimal number of moves it takes to solve that maze and the lateralization index as fixed effects, and participant IDs as random effects. We repeated this analyses to predict the deviation from the optimal path (i.e., the difference between the optimal number of moves and the total moves for a given trial).

We then explored whether participants were faster at navigating mazes in which the sVGC model more closely aligned with participants' awareness reports. To do so, we first regressed out the effects of the sVGC model predictions from the awareness reports of participants using a hierarchical linear regression model. We then took the mean squared residuals of each trial from this regression model and used that as a predictor in a second regression model. The mean squared residual of each trial represents the unexplained variance in participants awareness reports after accounting for the sVGC model, where larger numbers indicate more unexplained variance. Here, we predicted participants' response times from the optimal number of moves and the mean squared residuals as fixed effects, and participants' IDs as random effects.

## Data & Code Availability

All in-house code used for data analysis and visualization is available on GitHub https://github.com/jasondsc/ConsciousDetour. The reanalyzed data presented herein are available from https://www.nature.com/articles/s41586-022-04743-9. The data from experiment 3 are available from https://osf.io/sa6vf/.

## Acknowledgements

J.d.S.C. is supported by the Natural Sciences and Engineering Research Council of Canada (NSERC) postdoctoral fellowship program. S.M.F. is a CIFAR Fellow in the Brain, Mind and Consciousness Program and is supported by a Wellcome/Royal Society Sir Henry Dale Fellowship [206648/Z/17/Z] and UKRI under the UK government's Horizon Europe funding guarantee (selected as ERC Consolidator, grant number 101043666).